\documentstyle[11pt,paspconf,epsfig]{article}

\begin{document}

\title{Redshift Surveys and the Value of $\Omega$}

\author{Hume A. Feldman \& Adrian L. Melott}
\affil{Department of Physics \& Astronomy, University of Kansas,
Lawrence, KS 66045}

\begin{abstract}
We compare the statistical properties of structures normal and
transverse to the line of sight which appear in observational data from
redshift surveys. We present a statistic which can quantify this effect
in a conceptually different way from standard analyses of distortions of
the power-spectrum or correlation function. From tests with N--body
experiments, we argue that this statistic represents a new, more direct
and potentially powerful diagnostic of the cosmological density
parameter $\Omega$.
\end{abstract}



\section{Introduction}

There are multiple methods, both observational and numerical that
combine to constrain cosmological parameters. No single method is able
to determine by itself more than one of the main parameters with good
accuracy in a model--independent way. We introduce an approach which
bypasses the need to measure peculiar velocities or the underlying mass
distribution (bias) to probe $\Omega_m$ (the total matter density) by
directly examining displacements in redshift space. The method takes
advantage of the highly successful Zel'dovich approximation which
relates displacements to peculiar velocities in the weakly nonlinear
regime as a function of $\Omega$--plus the fact that peculiar velocities
look like displacements in redshift space.  This leads to a
bias--insensitive method to probe $\Omega$.

We explore a radically new way of estimating the mass density of the
Universe from redshift surveys. Unlike POTENT and power--spectral
redshift distortion methods, this method is insensitive to
bias. Furthermore, it does not depend on the expensive and
time--consuming measurement of peculiar velocities. It measures
$\Omega_m$ from redshift surveys directly in contrast to Supernova
projects which measure $q_0$, or CMB perturbations, which measure
complex combinations of model dependent parameters. This method is model
independent and is insensitive to the source of cosmic perturbations,
unlike the CMB power spectral methods, which only work when the
primordial power spectrum is known.

The formation of the largest structures in the universe ({\it i.e.}
galaxies, groups of galaxies, clusters, superclusters and voids of
galaxies) is a fascinating problem.  Many current questions ranging from
speculations on the physical nature of dark matter, to the measurement
of angular anisotropies of the microwave background radiation and
determination of the epoch of galaxy formation join together here.
These structures hold information about the very early stages of the
evolution of the universe.  This assumption is based on the fact that
the larger the object, the longer the characteristic time of its
evolution.  Thus, in terms of characteristic evolution time, the larger
the structure the younger it is.  Superclusters are dynamically
unrelaxed systems, and in studying them, one can learn about primordial
fluctuations in the universe.

We focus here on the {\it weakly nonlinear or quasi--linear regime}, in
terms of both dynamics and statistics.  The very largest scales are in
the linear regime.  They are observationally difficult to investigate,
but the dynamical questions are simple.  On the other hand, the deeply
nonlinear regime is difficult to connect with initial conditions. 

A new generation of the redshift surveys (SDSS, 2dF) will open up the
possibility of observational study the scales in the quasi-linear regime
($\sim100h^{-1}$Mpc, where $h$ is the Hubble constant in units of $100$
km/s/Mpc.) Also they will allow a much more detailed statistical
analysis of the structures on $30-100h^{-1}$Mpc scales. Studies of
geometry and topology of the largest structures which traditionally
suffered from small databases will play an important role in
discriminating cosmological scenarios.

One can characterize the weakly nonlinear regime as probing dense
concentrations ($\delta\rho/\rho \geq$ 1) which are still within reach
of Zel'dovich and other nonlinear approximations.  Very little if any
phase mixing or shell crossing has happened on these scales,
corresponding roughly to superclusters.  We think these scales deserve
more attention for two reasons: Observationally, new ground--based
redshift surveys are greatly increasing the quantity and quality of
data.  In terms of theory and analysis, new techniques show that it is
possible to make a direct link between this scale and initial
conditions.  Nearly all structure formation work in cosmology has either
focused on very large scales using linear theory or else galaxy/cluster
formation using hydrodynamics.  Superclusters provide information not
easily accessible to either approach.

\begin{figure}
{\psfig{file=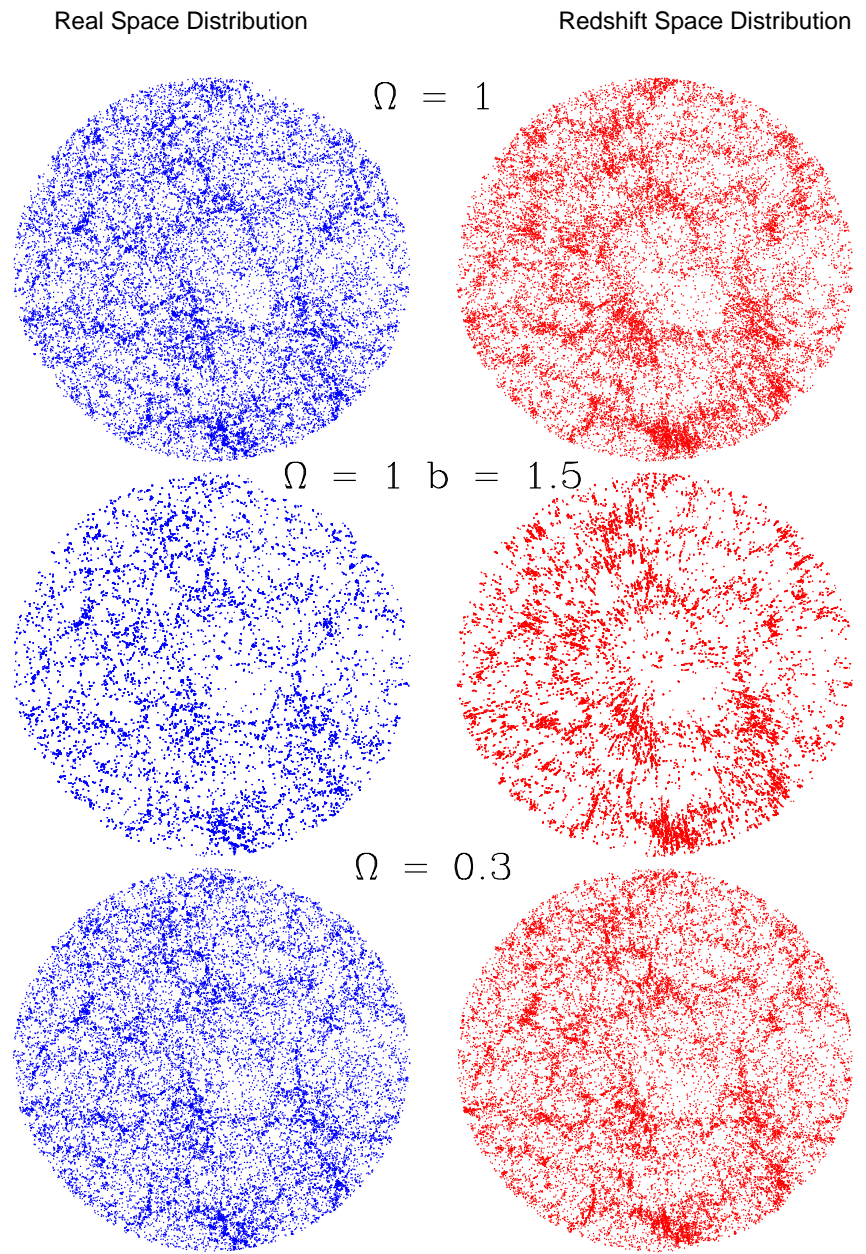,width=14cm}}
\caption{Illustration of redshift distortions for various models} \label{fig-1}
\end{figure}

\section{Method}

According to the Hubble law, $v\simeq cz \simeq H_0 r$, the recession
velocity $v$ of a galaxy, inferred from its redshift, is proportional to
its proper distance from the observer, $r$, $H_0$ is the Hubble constant
today. Irregularities gravitationally generate peculiar velocities
($v_p$), so that the true relationship is
\begin{equation}
v=H_0 r + v_p 
\label{hubble}
\end{equation}
where $v_p$ is the line of sight component of the peculiar motion. Maps
of galaxy positions constructed by assuming that velocities are exactly
proportional to distance (redshift space) have two principal
distortions. The first is generated in dense collapsed structures where
there are very many galaxies at essentially the same distance from the
observer, each with a random peculiar motion.  This results in a radial
stretching of the structure known as a ``Fingers of God''. The second
effect acts on much larger scales (e.g. Kaiser 1987).  A large
overdensity generates coherent bulk motions in the galaxy distribution
as it collapses. Material generally flows towards the center of the
structure, i.e. towards the observer for material on the far side of the
structure and away from the observer on the near side so it will appear
compressed along the line of sight. These effects are large for critical
$\Omega_m$ and negligible for small $\Omega_m$. (For illustration of
this effect see http://kusmos.phsx.ukans.edu/~feldman/redshift-distortions.html).

While redshift-space distortions are a nuisance when one wants to
construct accurate maps of the (true) spatial distribution of galaxies,
they may lead to a robust determination of $\Omega_m$ the contribution
of clustered matter, (baryonic or not) to the mass density of the
Universe.

It has been noticed that superclusters appear to
``surround" us, in a preferentially concentric pattern. Although the
statistics are poor (few superclusters), it is interesting to ask
whether this effect could appear in a homogeneous, isotropic
Universe. The effect of peculiar velocity breaks the isotropy in
redshift--space diagrams, interacting with inhomogeneities differently
depending on how their long axis is oriented.

Our new method is based on the suggestion in Melott et al. (1998); see
also Praton et al. (1997).  The essence of our method can be explained
based on the images in Figure 1 which are slices of 3D simulations.  The
left side is real--space, the right side redshift--space. The upper row
are evolved in a $\Omega_m=1$ cosmology ($\lambda=0$), the second row
are of a critical CDM cosmology with high bias, the lower an
$\Omega_m=0.1$ cosmology. All have similar large--scale linear power
amplitude and phases at the moment shown. It is clear that the visual
concentric effect are much stronger in the high $\Omega$ models in
redshift space.

The models we use to illustrate the method have the same initial power
spectrum (an $\Omega_m=1$ CDM model, normalized to a circle radius of
230 h$^{-1}$ Mpc for $h=0.67$ $\sigma_8=1$). We use the same spectrum
also for the low $\Omega_m$ case to make it clear that the effect is
rooted in $\Omega_m$, not the spectrum. All the slices have very nearly
the same number of particles.  The effect of peculiar motions is to
increase the spacing of large--scale structures in the radial direction
as compared with real space, as we show quantitatively below. The
enhancement is $\Omega_m$--dependent.  It is on the basis of this
visually--striking difference between the redshift--space behavior of
low-- and high--density models that we propose a statistic that
reproduces the eye's sensitivity to differences in pattern.
 
The essence of large--scale redshift space effects is a compression
and/or expansion effect along the line of sight.  It can best be
explained (following Melott et al, 1998) using the Zel'dovich
approximation (Zel'dovich 1970) This approximation follows the
development of structure by relating the final (Eulerian) position of a
particle ${\bf r}$ at some time $t$ to its initial (Lagrangian) position
${\bf q}$ defined at the primordial epoch when particles were smoothly
distributed:
\begin{equation} 
{\bf r}= a(t) {\bf x}({\bf q},t)= a(t)[{\bf q} - D_+ \nabla_{\bf q}
\Phi ({\bf q})].
\label{rzeld}
\end{equation} 
In this simple, separable mapping, the displacement field is given by
the gradient of the primordial gravitational potential $\Phi$, with
respect to the initial coordinates. $a(t)$ is the cosmic scale
factor. Differentiating this expression leads to
\begin{equation}
{\bf V}= {d {\bf r}\over dt} = H{\bf r}-a(t)\dot{D}_+ \nabla_{\bf q}
\Phi({\bf q}) 
\label{vzeld}
\end{equation} 
for the velocity of a fluid element ${\bf V}$, where $D_+$ is the
linear growth of perturbations as a function of time, usually
parameterized by $f=d\log D_+/d\log a$.  This is now known to reproduce
weakly non-linear (i.e. large--scale) features in the distribution of
matter very accurately indeed, if implemented in an optimized form known
as the Truncated Zel'dovich Approximation (Coles et al. 1993, Melott
1994). 

\begin{figure}
\centerline{\psfig{file=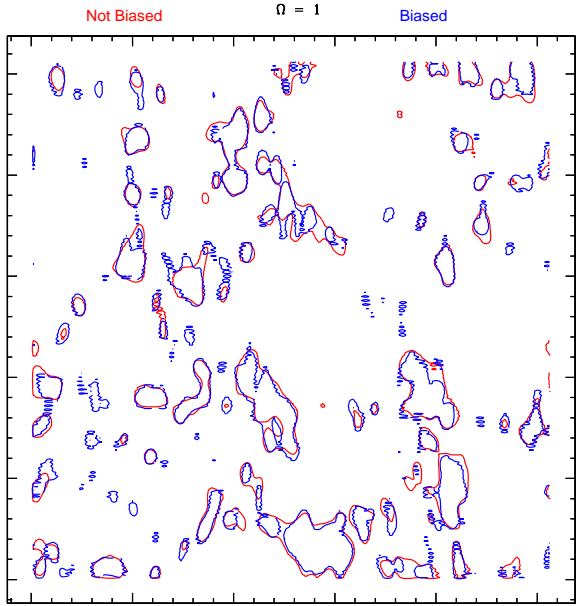,width=9cm}}
\caption{$\Omega=1$ biased (dotted lines) and non--biased (solid)
contour levels. }
\label{fig-2}
\end{figure}

\begin{figure}[t]
\centerline{\psfig{file=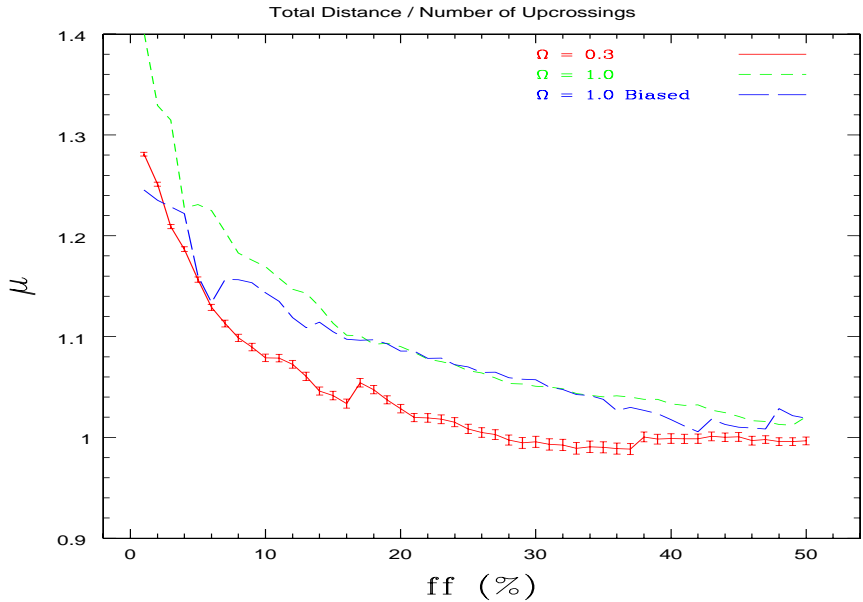,width=12cm,height=8cm}}
\caption{The ratio of parallel to perpendicular distances in a
redshift maps. $\mu$ is large for critical $\Omega$ models.} 
\label{fig-3}
\end{figure}

The mapping (\ref{vzeld}) provides a straightforward explanation of the changed
characteristic scale of structures in the redshift direction.
Calculating the redshift coordinate exactly and translating it into an
effective distance $d_z$ gives 
\begin{equation}
d_z={V\over H}=r_3 - f a(t)D_+(t)
\nabla_3 \Phi ({\bf q})=aq_3-(1+f) a(t)D_+(t) \nabla_3 \Phi ({\bf q}),
\label{zzeld}
\end{equation} 
in which we have the 3-axis in the redshift direction.  Thus the
displacement term becomes multiplied by a factor $(1+f)$ in
(\ref{zzeld}) compared to (\ref{vzeld}).  The effect of the displacement
field in redshift space is to give the observer a ``preview'' (albeit in
only one direction) of a later stage of the clustering hierarchy. (Note
that $\delta$, the density contrast, does not enter here).

We construct density contours for the smoothed field, and take
lines-of-sight through the smoothed density field and calculate the {\it
rms} distance between successive same-direction contour-(up)crossings of
high density levels; denoted $ S_{\parallel}$.  We also do a similar
calculation for lines in the direction orthogonal to the observer's line
of sight; denoted $ S_{\perp} $ After much experimentation with a large
ensemble of simulations, a simple statistic turned out to be nearly
optimal; the ratio of the {\it rms} spacing in the redshift direction to
that in the orthogonal direction, which we call $\mu$: 
\begin{equation}
\mu ={ S_{\parallel}\over S_{\perp}}
\label{mu}
\end{equation} 

In order to examine a density field (the galaxy density in redshift
surveys) it is necessary to specify a scale on which the density field
will be smoothed. In our case, we want the smoothing scale to include
the large--scale dynamics while filtering out small, fully nonlinear
dynamics. Then, we must chose one contour level to use for the
upcrossing interval measurement.

Although in general contours corresponding to a given filling factor
reduce bias dependence, a particular level must be chosen. We choose
that level corresponding to filled fraction of 1/8. This fraction has
the motivation that dissipationless collapse of a uniform medium will
virialize at about this volume fraction. Since with our choice of
smoothing we are looking only at just--collapsed structure, this is an
appropriate estimate. The use of a small fraction emphasizes the
interval between objects, not the size of the objects themselves. We
have checked that our measure also has a broad maximum around our choice
of filling factor, so that it is not especially sensitive to this
choice.

The fundamental object used in this method is the isodensity contour
level. Our steps are (1) Make a 2d array consisting of projections of a
slice of a 3--D distribution (2) Construct a smoothed density field (3)
Make isodensity contour levels corresponding to a set filling factor
i.e. fraction of the available area (4) Measure the distance between
upcrossings of this contour in the redshift and transverse (real)
directions.

To summarize, we show that the typical origin of bias-dependence is
absent; we argue that our filling factor approach eliminates another
possible source of bias; and we show results of a simulation which
behaves in this way (bias insensitive). As can be seen in Figure 2, the
bias dependence is negligible since the contours change little in the
biased model.

In figure 3 we show the results of the simulations. We plot $\mu$ (see
\ref{mu}) vs filling factor ($ff$). we added the error bars only to one
line, but they are of similar magnitude. We see that the critical
$\Omega$ models behave similarly and are significantly different than the
low ($\Omega=0.3$) model.

\section{Conclusions}

In the next few years, astronomers will map an appreciable fraction of
the Universe by redshift surveys where recession speed is assumed to be
proportional to distance. Gravity induce peculiar motions of
galaxies as part of the ongoing process of structure formation.  We
showed that such motions tend to enhance redshift structures concentric
about the observer, and argued that the strength of this effect may be a
powerful new probe of the mass density of the Universe.

\noindent The principal limitations of this method are the following:
\begin{itemize}
\item 
The smoothing length is specified by the autocorrelation
function. Biasing may affect this somewhat (small effect).
\item
The spacing ratio   $\mu\ne1$ for low $\Omega$
models. This is due to the fact that the ``fingers of God'' effect
introduce noise (small effect). 
\item
The method requires deep, dense, 3--D  redshift surveys where the
correlation length is much smaller than the survey effective
radius. These surveys are coming (SDSS, 2DF)
\item
The method measures $\Omega_m$ not $\Lambda$.
\end{itemize}

\noindent The advantages of this method are:

\begin{itemize}
\item 
No need for distance measurements, redshifts are enough.
\item
No comparison between the density field and the velocity field. Thus we
measure $\Omega$ directly no $\beta=\Omega^{0.6}/b$, that is, virtually
no bias dependence.
\item
Bias affects this statistic only through excess smoothing which is both
a weak effect and can be controlled easily.
\end{itemize}

\acknowledgments I would like to thank the conference organizers for a
fascinating and well--run meeting. This work was supported in
part by the NSF-EPSCoR program and the GRF at the University of Kansas.

\end{document}